# Two Dimensional Electron and Hole Gases at the Surface of Graphite

S.V. Morozov[1,2], K.S. Novoselov[1], D. Jiang[1], A. A. Firsov[2], S.V. Dubonos[2] & A.K. Geim[1]

[1]Department of Physics, University of Manchester, M13 9PL, Manchester, UK
[2]Institute for Microelectronics Technology, 142432 Chernogolovka, Russia

We report high-quality two-dimensional (2D) electron and hole gases induced at the surface of graphite by the electric field effect. The 2D carriers reside within a few near-surface atomic layers and exhibit mobilities up to 15,000 and 60,000 cm$^2$/Vs at room and liquid-helium temperatures, respectively. The mobilities imply ballistic transport at micron scale. Pronounced Shubnikov-de Haas oscillations reveal the existence of two types of carries in both 2D electron and hole gases.

{This paper summarises our knowledge about the electric field effect and electronic properties of relatively thick graphite films (thicker than of about 10 monolayers). Some preliminary data for such films were previously reported by us in cond-mat/0410631 and also included as a part of supplementary information in *Science* **306**, 666 (2004). The latter paper was dedicated to ultra-thin films (3 to 10 layers thick) which exhibited behaviour qualitatively different from the one of thicker films and described in the present manuscript.}

2D gases have proved to be one of the most pervasive and reach-in-phenomena systems and, deservedly, they have been attracting intense interest of physicists and engineers for several decades, leading to the discovery of a whole range of new applications and phenomena including the widely-used field-effect transistor and the integer and fractional quantum Hall effects. So far, all 2D systems (2DS) have been based on semiconducting materials where carriers are induced by either local doping or the electric field effect (EFE) [1]. As concerns metallic materials, many earlier efforts have proven it difficult to change intrinsic carrier concentrations by EFE even in semimetals (see, e.g., [2,3]), and a possibility of the formation of 2D gases in such materials was never discussed. The origin of these difficulties lies in the fact that charge densities induced by EFE cannot normally [4] exceed $\approx 10^{13} cm^{-2}$, which is several orders of magnitude smaller than area concentrations in a nm-thin film of a typical metal. Accordingly, possible EFE would be obscured by a massive contribution from bulk electrons. Prospects of the observation of a fully-developed 2DS in a metallic material seem to be even more remote, because locally-induced carriers could merge with the bulk Fermi sea without forming a distinct 2DS. Furthermore, because the screening length in metals never exceeds a few Å, EFE-induced carriers may also end up as a collection of puddles around surface irregularities rather than form a continuous 2DS.

In this Letter, we report a strong ambipolar field effect at the surface of graphite. We have investigated EFE-induced carriers in this semimetal by studying their Shubnikov-de Haas (SdH) oscillations and analyzing the oscillations' dependence on gate voltage $V_g$ and temperature $T$. This has allowed us to fully characterize the carriers and prove their 2D character. The 2D electron and hole gases (2DEG and 2DHG, respectively) exhibit a surprisingly long mean free path $l \approx 1 \mu m$, presumably due to the continuity and quality of the last few atomic layers at the surface of graphite where the 2D carriers are residing. Our results are particularly important in view of current intense interest in the properties of thin [5-9] and ultra-thin [10,11] graphitic films and recently renewed attention to anomalous transport in bulk graphite [12,13].

In our experiments, in order to minimize the bulk contribution, we used graphite films with thickness $d$ between 5 to 50 nm. They were prepared by micromechanical cleavage of highly-oriented pyrolytic graphite (HOPG) and placed on top of an oxidized Si wafer, as described in detail in [14]. Multi-terminal transistor-like devices were then fabricated from these films by using electron-beam lithography, dry etching and deposition of Au/Cr contacts [14]. Fig. 1 shows one of our experimental devices. We studied more than two dozen of such devices by using standard low-frequency lock-in techniques at $T$ between 0.3 and 300K in magnetic fields $B$ up to 12T. By applying a gate voltage between the Si wafer and graphite films, we could induce a surface charge density in graphite of $n = \varepsilon_0 \varepsilon V_g/te$, where $\varepsilon_0$ and $\varepsilon$ are the permittivities of free space and SiO$_2$, respectively, $e$ is the electron charge and $t = 300$ nm is the thickness of SiO$_2$. The above formula yields $n/V_g \approx 7.18 \cdot 10^{10}$ cm$^{-2}$/V and, for typical $V_g \approx 100$V, $n$ exceeds the intrinsic density $n_i$ of carriers per single layer of graphite by a factor of >20 (graphite has equal concentrations of holes and electrons, and $n_i \approx 3 \cdot 10^{11}$cm$^{-2}$ at 300K [15]). Because the screening length in graphite is only $\approx 0.5$nm [16] and the interlayer distance is 0.34nm, the induced charge is mainly located within one or two surface layers whereas the bulk of our graphite films (15 to 150 layers thick) remains unaffected. In a sense, the thickness of graphite is not important in our experiments but it has to be minimized to reduce parallel conduction through the bulk and allow accurate EFE measurements.

A typical behavior of conductivity $\sigma$ and Hall coefficient $R_H$ as a function of $V_g$ is shown in Figs. 1 and 2. Conductivity increases with increasing $V_g$ for both polarities, which results in a minimum close to zero $V_g$. The observed changes in $\sigma$ amount up to 300% for the 5nm film and can still be significant (>20%) even for $d \approx 50$nm (Fig. 1). As the polarity changes, $R_H$ sharply reverses its sign and, at high $V_g$, it decreases with increasing $V_g$ (Fig. 2). The observed behavior can be understood as due to additional near-surface electrons (holes) induced in graphite by positive (negative) $V_g$. Indeed, one can write $\sigma(V_g) = \sigma_B + n(V_g)e\mu$ where $\sigma_B$ is the bulk conductivity and the second term describes the EFE-induced conductivity. If $\mu$ is independent of $V_g$, then $\Delta\sigma = \sigma - \sigma_B \propto |V_g|$ which qualitatively explains the experimental behavior. As concerns the Hall effect, assuming for simplicity equal mobilities for all carriers, the standard two-band model [15] yields $R_H = (n_h - n_e)/e(n_h + n_e)^2$ where $n_h \approx n_e$ are the area concentrations of holes and electrons, respectively, including both bulk and EFE-induced carriers. This leads to $R_H \approx 1/ne \propto V_g^{-1}$, if $n$ is larger than bulk carrier concentrations, and $R_H \propto n \propto V_g$ at low $V_g$ (see Fig. 2). A full version of the above model (using mobilities as fitting parameters) allowed us to describe the observed $\sigma$ and $R_H$ for all gate voltages, similarly to the analysis given in [10,14]. For brevity, we do not include this numerical analysis in the present report. We also note that the discussed minimum in $\sigma$ was often found to be shifted from zero $V_g$ [14]. The sample-dependent shift occurred in both directions of $V_g$ and is due to unintentional chemical doping of graphite surfaces during microfabrication [10].

From the observed changes in $R_H = 1/ne$ at high $V_g$ we have calculated $n$ as a function of $V_g$ and found that changes in $n$ are accurately described by $n/V_g \approx 7.2 \cdot 10^{10}$ cm$^{-2}$/V, in agreement with the earlier formula. This proves that there are no trapped charges and all EFE-induced carriers are mobile. In addition, the linear dependence of σ on $n$ allowed us to find carriers' mobilities as $\mu = \sigma/ne$. The mobilities varied from sample to sample between 5,000 and 15,000 cm$^2$/Vs at 300K, reaching up to 60,000 cm$^2$/Vs at 4K in some devices. Thicker films generally exhibited higher μ. For a typical $n \approx 10^{13}$cm$^{-2}$, the above mobilities imply $l \approx 0.5$ and 2 μm at 300 and 4 K, respectively. For comparison, macroscopic samples of our HOPG exhibited $\mu \approx 15,000$ cm$^2$/Vs at 300K and >100,000 cm$^2$/Vs at 4 K.

To characterize the near-surface carriers further, we studied magnetoresistance $\rho_{xx}$ of our devices at liquid-helium $T$. Fig. 3 shows a typical behavior of $\rho_{xx}(B)$. There is a strong linear increase in $\rho_{xx}(B)$, on top of which SdH oscillations are clearly seen. Below we skip discussions of the linear magnetoresistance (we believe it is due to the so-called parallel conductance effect, where the electric current redistributes with increasing $B$ being attracted to regions with lower μ) and concentrate on the observed oscillations. Our devices generally exhibit two types of SdH oscillations, dependent and independent of $V_g$. The latter are more pronounced in thicker devices and attributed to the bulk unaffected by EFE. On the other hand, the oscillations dependent on $V_g$ indicate near-surface carriers and were dominant in thinner samples. The latter oscillations exhibited a clear 2D behavior discussed below.

First, we carried out the standard test for a 2DS by measuring SdH oscillations at various angles θ between $B$ and graphite films. The oscillations were found to depend only on the perpendicular component of magnetic field $B\cos\theta$, as expected for 2D carriers. This test is however not definitive, as the cosθ-dependence was also observed in bulk HOPG because of its elongated Fermi surface [15,17]. Therefore, in order to identify dimensionality of the field-induced carriers, we have used another test based on the fact that different dimensionalities result in different dependences of the Fermi energy as a function of $n$, and the measured frequency of SdH oscillations $B_F$ should vary as $\propto n$ or $\propto n^{2/3}$ for 2D and 3D cases, respectively [18,19]. Rather unusually for 2DS [1], accurate measurements of $B_F(n)$ were possible in our case.

Fig. 3 shows examples of changes in frequency of SdH oscillations with varying $V_g$ and their analysis based on the standard Landau fan diagrams. Although time consuming, such analysis is most reliable, when there is a limited number of oscillations. The observed minima can be separated into different sets of the SdH frequencies, indicating different types of carriers characterized by different $B_F$ (note that $B_F$ is also the field corresponding to a filling factor $N =1$). Importantly, we found that the phase of the SdH oscillations was always the same as in the standard 2DEG (namely, minima in $\rho_{xx}$ in high $B$ occur at integer $N$; see inset in Fig. 3). According to [13,19], this phase indicates carriers with a finite mass $m$ [18].

Analysis as in Fig. 3 was carried out for many gate voltages and several samples. Our results are summarised in Fig. 4, which shows $B_F$ as a function of $n$ observed in five different devices. One can clearly see 4 sets of SdH frequencies, two for each gate polarity, indicating light and heavy electrons and holes induced by EFE. For clarity, SdH frequencies due to bulk carriers are omitted (two sets of such gate-independent $B_F$ were normally observed in thicker devices). The first important feature of the shown curves is the fact that $B_F$ depends linearly on $n$. The dependence $B_F \propto n^{2/3}$ expected for 3D carriers as well as for carriers in bulk graphite [15] cannot possibly fit our data. This unequivocally proves the 2D nature of the field-induced carriers at the surface of graphite.

Our data in Fig. 4 also show that the observed light and heavy 2D carriers account for the entire charge $n$ induced by EFE [20]. Indeed, we can write $n = n^h + n^l$, or $n = (2e/h) \cdot g^h B_F^h + (2e/h) \cdot g^l B_F^l$ where upper indices $h$ and $l$ refer to heavy and light carriers, respectively, and $g$ is their valley degeneracy. Taking into account that $B_F = \alpha \cdot n$, the above expression can be re-written as $(2e/h) \cdot g^h \alpha^h + (2e/h) \cdot g^l \alpha^l = 1$. For our 2DEG, the best fits in Fig. 4 yield $\alpha^l \approx 1.75 \cdot 10^{-12}$ Tcm$^2$ and $\alpha^h \approx 6.7 \cdot 10^{-12}$ Tcm$^2$, which leads to the numerical equation $0.085\{4\%\}g^l + 0.325\{2\%\}g^h = 1$ where $\{\%\}$ indicates the coefficients' accuracy. As $g^{h,l}$ have to be integer, the equation provides a unique solution with $g^h = 2$ and $g^l = 4$. No other solution is possible. Similar analysis for the 2DHG yields $\alpha^l \approx 3.7\{10\%\}$ and $\alpha^h \approx 6.7\{5\%\}$ in units of $10^{-12}$ Tcm$^2$, which again provides only one solution $g^l = g^h = 2$. Note that all our samples showed exactly the same 2D electron behavior. The situation for 2D holes is more complicated as in some samples we also observed slopes $\alpha^l \approx 1.4\{10\%\} \cdot 10^{-12}$ Tcm$^2$ and $\alpha^h \approx 8.9\{5\%\} \cdot 10^{-12}$ Tcm$^2$ ($g^l = g^h = 2$). The origin of the different behaviors remains unclear.

To find the masses of EFE-induced carriers, we have measured SdH oscillations' amplitude $\Delta\rho$ as a function of $T$ at high $n \approx 10^{13}$cm$^{-2}$ where the oscillations due to heavy carriers were best resolved. For heavy 2D electrons, the fit by the standard expression $T/\sinh(2\pi^2 k_B Tm/\hbar eB)$ yields $m_e^h = 0.06 \pm 0.05 m_0$ (see Fig. 4). Similarly, for heavy 2D holes we obtained $m_h^h = 0.09 \pm 0.01 m_0$. Masses of light carriers were then found as follows. If the gate voltage changes by $dV_g$, the Fermi energy has to shift by an equal amount for both light and heavy carriers. These leads to the expression $(dB_F/dn)^l/m^l = (dB_F/dn)^h/m^h$, which shows that the ratio $\alpha^h/\alpha^l$ yields the ratio between heavy and light masses. In the case of our 2DEG, we obtain $m_e^l \approx 0.015 m_0$, while for the 2DHG in Fig. 4 $m_h^l \approx 0.05 m_0$. For comparison, in bulk graphite one usually finds two types of holes and only one type of electrons with $m_e^h \approx 0.056 m_0$ [15,21] $m_h^h \approx 0.084 m_0$ [15] or $\approx 0.04 m_0$ [21] and $m_h^l \approx 0.003 m_0$ [22]. Theory expects heavy carriers to have $g = 2$, whereas the location and degeneracy of minority holes are uncertain even for bulk graphite, being sensitive to, e.g., minor changes in the interlayer spacing. The existence of two electron carriers (one with $g = 4$) and two types of relatively heavy holes clearly distinguish between bulk and surface carriers in graphite. Also, our 2D carriers are clearly different from those reported for ultra-thin graphite films [10] and expected for a graphite monolayer [15]. It requires dedicated band-structure calculations to understand these differences and the nature of the observed carriers.

In conclusion, we have presented a comprehensive description of 2D electron and hole gases formed at the surface of graphite by electric field effect. This is the first non-semiconducting 2D system, which also stands out from the known 2D gases due to its extremely narrow quantum well, strong screening by bulk electrons, highly mobile carriers located directly at the surface and an unusual layered crystal structure of the underlying material.

[17] It requires measurements at $\theta > 85°$ [15] to distinguish between such elongated Fermi surfaces and a true 2DS. No SdH oscillations survived in our devices for such shallow angles.

[18] A single layer of graphite is expected to be a zero-gap semiconductor with a linear dispersion spectrum and massless (Dirac) carriers [15]. As the EFE-induced carriers are mainly located within one or two near-surface layers, one might also expect the carriers to be massless. No evidence for the latter was found in the experiments, whereas the observed phase of SdH oscillations seems to indicate the opposite [13,19]. Accordingly, we assume normal, massive carriers in this report. We note however that except for the phase our other results cannot distinguish between massive and massless carriers. For example, for 2D Dirac fermions, $B_F$ is also a linear function of $n$, whereas the masses extracted from $T$-dependence of SdH oscillations could then be interpreted as "cyclotron masses" of Dirac fermions [19].

[20] Results of Fig. 4 also indicate that only one spatially quantized 2D subband is occupied. Indeed, if the second subband were to become populated at some $V_g$, this would result in a drastic change in slopes on the $B_F(n)$-curves.

FIGURE CAPTIONS

Figure 1. Electric field effect in graphite. Conductivity σ as a function of gate voltage $V_g$ for graphite films with $d \approx 5$ and 50 nm (main panel and the upper inset, respectively); $T = 300$K. For the 5nm device, $\mu \approx 11,000$ and 8,500 cm$^2$/Vs for electrons and holes, respectively. Left inset: schematic view of our experimental devices. Right inset: optical photograph of one of them ($d \approx 5$ nm; the horizontal wire has a 5 μm width).

Figure 2. Hall coefficient $R_H$ as a function of $V_g$ for the 5 nm device of Fig. 1. Inset: Hall resistivity $\rho_{xy}(B)$ for various gate voltages. From top to bottom, the plotted curves correspond to $V_g = -30, -100, -2, 100$ and 20V. Close to zero $V_g$, $\rho_{xy}$-curves are practically flat indicating a compensated semimetal whereas negative (positive) $V_g$ induce a large positive (negative) Hall effect. Solid curves in the main inset show the dependences $R_H \propto n$ and $R_H = 1/ne$ expected at low and high $V_g$, respectively.

Figure 3. SdH oscillations in a 5-nm film at three gate voltages (main panel). Note that the frequency of SdH oscillations increases with increasing $V_g$ (concentration of 2D electrons increases). The lower panel magnifies the oscillations for one of the voltages ($V_g = 90$V) after subtracting a linear-magnetoresistance background. The inset shows an example of the SdH fan diagrams used in our analysis to find SdH frequencies. $N$ is the number associated with different oscillations' minima.

Figure 4. SdH frequencies $B_F$ as a function of carrier concentration $n$. Different symbols indicate oscillations due to near-surface carriers in different devices. The data for different samples were aligned along the x-axis so that zero $n$ corresponded to zero $R_H$ and minimum σ, which allowed us to take into account the chemical shift [10] observed in some films. Solid lines are the best linear fits. The inset shows amplitude Δρ of SdH oscillations as a function of $T$ for 2D electrons and holes (open and solid symbols) at $B_F = 85$ and 55T, respectively. Solid curves are the best fits allowing us to find the carriers' cyclotron masses.

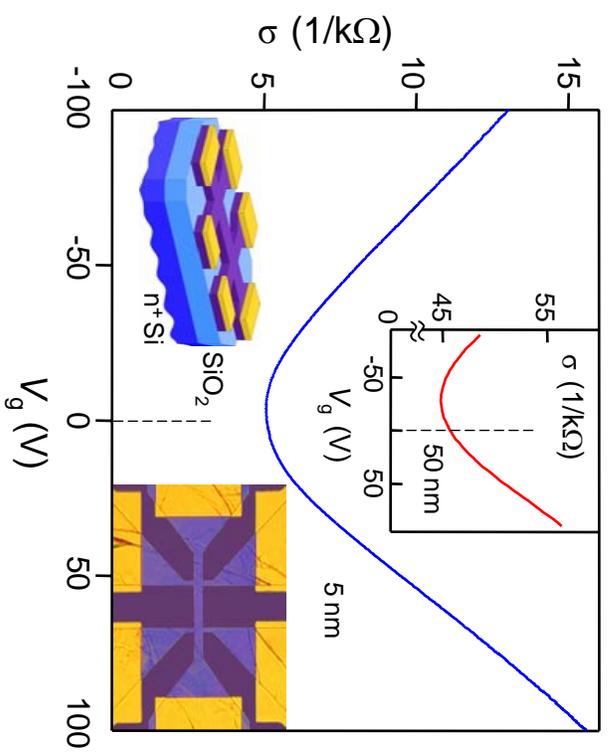

Figure 1

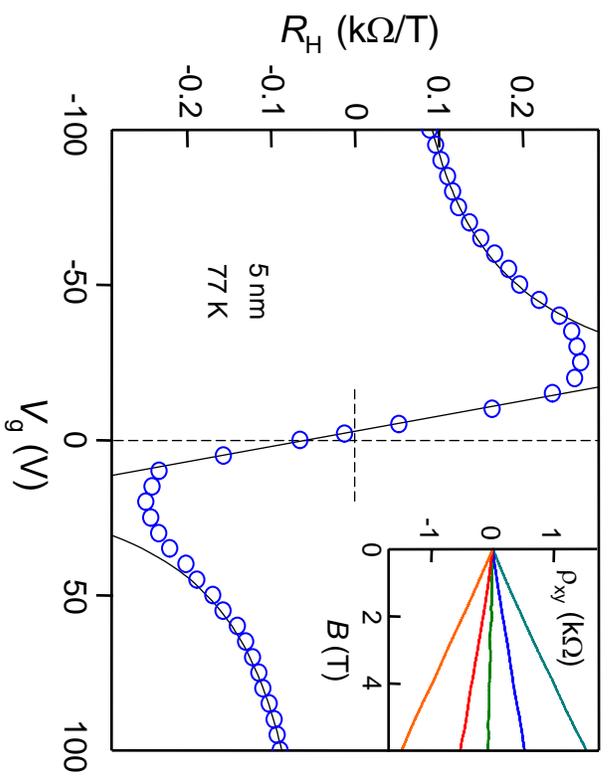

Figure 2

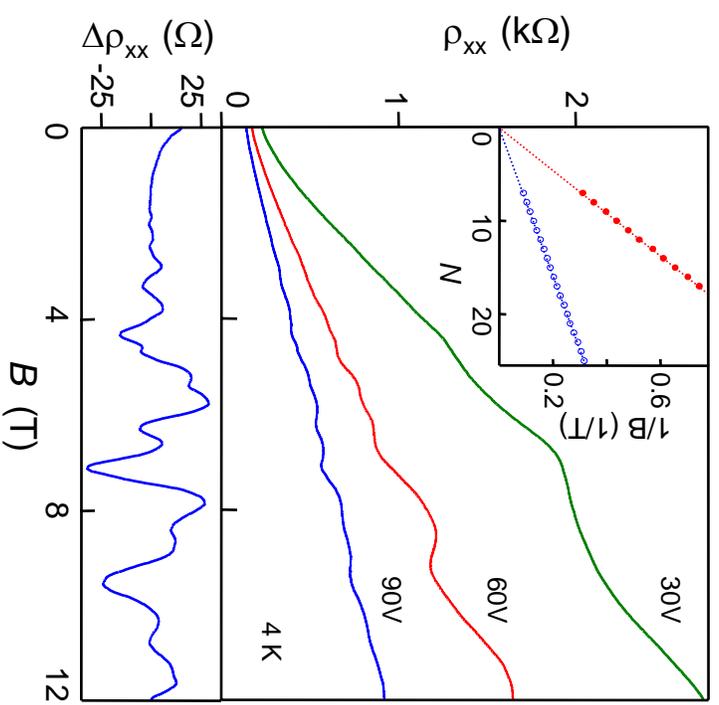

Figure 3

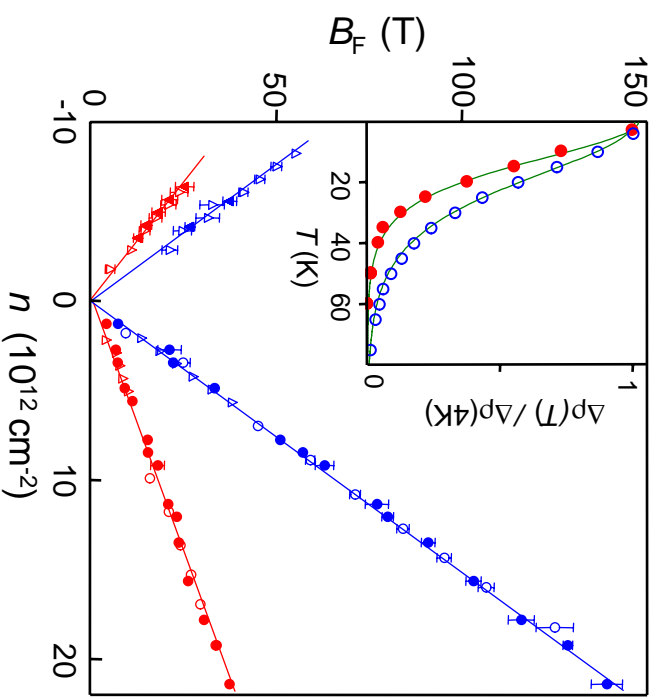

Figure 4